# High magnetic field scales and critical currents in SmFeAs(O,F) crystals: promising for applications

Philip J.W. Moll[1*], Roman Puzniak[2], Fedor Balakirev[3], Krzysztof Rogacki[4], Janusz Karpinski[1], Nikolai D. Zhigadlo[1], and Bertram Batlogg[1]

[1]Laboratory for Solid State Physics, ETH Zurich, Schafmattstr.16, CH-8093 Zurich, Switzerland,

[2]Institute of Physics, Polish Academy of Sciences, Aleja Lotników 32/46, PL-02-668 Warsaw, Poland, [3]

National High Magnetic Field Laboratory, Los Alamos National Laboratory, Los Alamos, New Mexico

87545, USA, [4]Institute of Low Temperature and Structure Research, Polish Academy of Sciences, ul.

Okolna 2, PL-50-422 Wroclaw, Poland

Correspondence should be addressed to P.J.W.M (phmoll@phys.ethz.ch)

**With the discovery of new superconducting materials, such as the iron pnictides[1], exploring their potential for applications is one of the foremost tasks. Even if the critical temperature $T_c$ is high, intrinsic electronic properties might render applications rather difficult, particularly if extreme electronic anisotropy prevents effective pinning of vortices and thus severely limits the critical current density, a problem well known for cuprates[2-5]. While many questions concerning microscopic electronic properties of the iron pnictides have been successfully addressed[5] and estimates point to a very high upper critical field[6-9], their application potential is less clarified. Thus we focus here on the critical currents, their anisotropy and the onset of electrical dissipation in high magnetic fields up to 65 T. Our detailed study of the transport properties of $SmFeAsO_{0.7}F_{0.25}$ single crystals reveals a promising combination of high ($>2x10^6 A/cm^2$) and nearly isotropic critical current densities along all crystal directions. This favorable intragrain current transport in SmFeAs(O,F), which shows the highest $T_c$ of 54 K at ambient pressure[10-12], is a crucial requirement for possible applications. Essential in these experiments are 4-probe measurements on Focused Ion Beam (FIB) cut single crystals with sub-$\mu m^2$ cross-section, with current along and perpendicular to the crystallographic $c$-axis.**

$SmFeAsO_{0.7}F_{0.25}$ single crystals - as all members of the so called "1111" family LnFePn(O,F) (Ln: Lanthanide, Pn: Pnictogen) - grow as thin platelets up to 200 x 200 x 10 $\mu m^3$ from NaCl/KCl flux under high pressure (30 kbar)[13,14]. Reliable measurements perpendicular to the plates, i.e. along the crystallographic $c$-direction, require new methods of sample preparation. Here we have employed a Focused Ion Beam (FIB) as a powerful tool for shaping and contacting on sub-$\mu m$ dimensions. A $Ga^{2+}$ ion beam is accelerated by 30 kV and focused onto the target, where it ablates the crystal without collateral damage. In addition to ablation, the FIB is also used to deposit $\mu m$-sized electric leads by ion-assisted chemical vapor deposition of Pt contained in a precursor gas, without exposing the structure to air. We have confirmed that FIB processing does not damage the crystals by comparing samples prepared with and without the FIB. This method of contacting, without previous cutting, was successfully applied to pnictide single crystals for in-plane currents.[14,15] We have used the FIB to cut into the same single crystal three contiguous 4-probe resistance bars, one in the $ab$-plane and two along the $c$-axis. The typical cross-section of the bars is ~ 1 μm$^2$ and the voltage probes are 5 μm apart on the $c$-axis bars and 35 μm on the bar along the $ab$-plane (Figure 1, for more details see appendix E). The resulting large geometry factor Length / Cross-section of the resistance bars of ~ 5-35/μm is appropriate for a four-point geometry. This U-shaped structure is optimized for the simultaneous measurement of $c$-axis and $ab$-plane resistivity on the same small section of a single crystal under identical conditions and is therefore well suited to address questions of electronic anisotropy in detail.

The temperature dependence of the $ab$- and $c$-direction resistivity in the normal state in zero field is given in Figure 2. The $ab$-plane resistance decreases almost linearly down to $T_c$ (~48 K) with $\rho(300\ \text{K})/\rho(50\ \text{K}) = 3.3$, in good agreement with previous studies on '1111' pnictides[11,14-16]. Along the $c$-axis, the resistivity continuously increases superlinearly from 300 K to $T_c$ and the anisotropy $\rho_c/\rho_{ab}$ increases from 2 at 300 K to 10 at $T_c$, following a particularly simple T dependence, parameterized by $d\ln(\rho_c/\rho_{ab})/dT = -1/T_0$ with $T_0 = 86.4$ (±0.6) K. We found the anisotropy to continue this trend at lower temperatures by suppressing superconductivity in pulsed magnetic fields (dots in Fig. 2), similar to the qualitative behavior in e.g. $Bi_2Sr_2CuO_y$[17]. To our knowledge, this is the first measurement of $\rho_c$ in the '1111' compounds, and its temperature dependence is strikingly different from that in '122' crystals ($Ba(Fe,Co)_2As_2$[18], $KFe_2As_2$[19]), where a decrease of $\rho_c$ is found. Just above $T_c$ the resistivity is 0.25 mΩcm along the $ab$-plane and 2.65 mΩcm parallel to the $c$-axis, yielding a resistivity anisotropy $\rho_c/\rho_{ab} = 10.6$. This ratio is low compared to most cuprates (e.g. ~ 80 in $YBa_2Cu_3O_{7-x}$[20], > 1000 for $Bi_2Sr_{2-x}La_xCuO_{6+\delta}$[21]), but higher than in $Ba(Fe,Co)_2As_2$ (~4)[18].

The U-shaped samples are very well suited for studies in pulsed magnetic fields because both current directions are simultaneously probed and the large geometry factor and sample resistance result in an excellent SNR (>110Ω in 4-probe above $T_c$). The high field experiments were performed at the National High Magnetic Field Laboratory (LANL/NHMFL) in pulsed fields up to 65 T (100 ms pulse duration). A 100 μA excitation current was passed through the sample and the voltage signals on all legs were recorded simultaneously using a custom digital lock-in amplifier operating at 64 kHz. The average noise floor in pulsed fields corresponds to < 0.25% $\rho_n$ for j||$ab$ and < 0.05% $\rho_n$ for j||$c$, reflecting the larger normal state resistance along the $c$-axis.

In Figure 3, the magnetoresistance of $SmFeAsO_{0.7}F_{0.25}$ is shown for field orientations parallel and perpendicular to the $c$-axis and for current along both orientations. The four panels capture the main qualitative results: The over-all feature and field-scale of the resistive transition is mainly determined by the orientation of the magnetic field with respect to the crystal axes, and is essentially independent of the current direction. This is an important prerequisite for the application in superconducting wires as the supercurrent in a polycrystalline wire is always limited by the least favorably oriented crystallites. We note the remarkable observation that the behavior in the only Lorentz force free configuration H∥$c$, j∥$c$ (lower left panel) is not significantly different from those with present Lorentz force at this level of dissipation. At a given temperature, the low dissipation region ('$\rho$=0') extends to higher magnetic fields when the field is parallel to $ab$ (Right panels of Figure 3). As the resistance in the superconducting state is caused by vortex motion, these results indicate that vortices are more strongly pinned if they are parallel to the plane. This dependence of the resistivity on the orientation of the magnetic field is typical for layered superconductors, but is least pronounced in the pnictides.[14-15,22-23]

To identify the field scales for potential applications, we estimate the magnetic field H* up to which dissipation-free current transport can be maintained. Given the noise floor in pulsed fields, we have adopted a conservative definition of H*(T) at $\rho(T,H^*) = 10$ $\mu\Omega$cm. This low level of dissipation, however, is still significantly larger than technical definitions of dissipation free transport ($\rho < 10^{-6}$ $\mu\Omega$cm). We have also measured $\rho$ near $T_c$ down to $10^{-3}$ $\mu\Omega$cm in dc-fields up to 14 T with significantly lower noise. When H*(T) is defined at lower levels of dissipation, it is shifted to lower temperatures by ~ 0.5 K per decade of $\rho$. The thus defined field scale H* is extremely high for H∥$ab$, i.e. > 50 T already 10 K below $T_c$ of 48 K, and even for H∥$c$, H* extrapolates well above 50 T at temperatures below 15 K. Large values of H* are expected from previous estimates for '$H_{c2}$', since H* defined at lower levels of dissipation is linked to '$H_{c2}$' non-trivially via vortex dynamics (an analysis of commonly used criteria for '$H_{c2}$' is given in the appendix A). Even more relevant for technical aspects, however, is the weak dependence of H* on the current direction and thus an additional important requirement for operation in high fields is met.

For applications in superconducting magnets, large critical current densities $j_c$ are also required. In the following discussion of the experimental results it is worth keeping in mind that the value of "$j_c$" depends on the dissipation level inherent to a particular experiment. We thus adapt the notation $j_c^{trans}$ for the critical currents measured by direct transport and $j_c^{mag}$ for magnetically measured critical currents. We found a nearly-isotropic intragrain critical current density in the range of $10^6$ A/cm$^2$, employing the following two methods: The most direct way to obtain $j_c^{trans}$ is to increase the applied current through a superconducting sample until a potential drop can be observed. Such experiments require extremely large currents for usual cross-sections, and thereby endanger the sample and cause significant heating at the contacts. We have employed the FIB again to carve a structure featuring two free-standing nanoscopic bridges out of the same SmFeAsO$_{0.7}$F$_{0.25}$ single crystal lamella, one bridge along the *c*-axis and one in the *ab*-plane (Figure 4a, cross-section: 608 nm x 768 nm, length 800 nm (*c*-axis), 566 nm x 860 nm, length 2 μm (*ab*-plane), ~ 1 Ω contact resistance). This structure enabled us to directly measure current densities up to 1.6 x 10$^6$ A/cm$^2$ without exceeding 7 mA. The critical transport experiments were performed in a Quantum Design PPMS 14 T superconducting magnet using a Keithley Delta Pulse System. Rectangular 123 μs current pulses of increasing amplitude were passed through the sample at 1s intervals to ensure proper thermalization and the voltage drop across the bridge was integrated for 68 μs (after 55 μs risetime). To avoid damage to the sample, the pulse ramp was stopped when a voltage limit of 15 μV was reached and a criterion of 5 μV was used to define j$_c$$^{trans}$ (Inset in Figure 4b shows raw I-V data). The shape and width of the onset of observable resistivity did not change when we increased the pulse duration by a factor of 3, and thus self heating could be ruled out in good agreement with numerical thermal simulations. To further prove the high reproducibility of our results, these measurements were performed on 3 different samples using different pulsed and continuous current measurement methods, which all gave very similar results (Appendix C). Even as the bridge geometry was not optimized to fully eliminate effects of corners, these results show that transport of current with densities of at least 1.6 x 10$^6$ A/cm$^2$ can be achieved at low temperatures.

The second method, used to estimate j$_c$$^{mag}$, involves measurements of the magnetization loops M(H) of macroscopic crystals, oriented with the magnetic field either perpendicular or parallel to the *ab*-planes. These measurements were performed in a SQUID magnetometer ("Quantum Design MPMS"). From the width ΔM(H) of the loops j$_c$$^{mag}$ has been calculated using Bean's model[24,25], which has been shown to be a realistic description of the current configuration also in Ba(Fe,Co)$_2$As$_2$[26].

The results for j$_c$$^{trans}$ and j$_c$$^{mag}$ are given in Figure 4, and each of them emphasizes different aspects. Starting at low temperature, particularly noteworthy are: (1) j$_c$$^{trans}$ and j$_c$$^{mag}$ exceed 10$^6$ A/cm$^2$, (2) are essentially independent of the field and only weakly field and current orientation dependent up to 14 T, and (3) the two methods applied to different crystals from different growth runs yield very similar results. Remarkably, the absolute values are very similar in all experiments, considering the use of different samples, slightly different doping concentrations during sample growth and the error resulting from the estimation of the effective crystal volume for the magnetization measurements.

The large values of the critical current reflect the highly effective pinning in this materials class[27] and give reason to expect that it can be further enhanced by proper material design and treatment. Even at the present level, j$_c$$^{trans}$ and j$_c$$^{mag}$ reach values that are generally considered a minimal requirement for technical applications[28]. Within the experimental accuracy, both $j_c^{trans}$ and $j_c^{mag}$ are independent of the magnetic field orientation at low temperatures. Furthermore, the transport measurements directly show that the critical current density $j_c^{trans} \| c$ along the $c$-axis is very similar to the $ab$-plane value at lowest temperatures, giving a critical current ratio of $j_c^{trans} \| c / j_c^{trans} \| ab \sim 2.1$. In Ba(Fe,Co)$_2$As$_2$ crystals (T$_c \sim 23$ K) with critical currents in the range of $10^5$ A/cm$^2$, the corresponding ratio is $\sim 1.5 - 3$[18,26]. At higher temperatures and in high magnetic fields, $j_c^{trans} \| c$ decreases more rapidly than $j_c^{trans} \| ab$ but stays above the high level of $10^5$ A/cm$^2$ in fields up to 14 T and temperatures up to 25 K. It remains an open question whether the origin of the pronounced upturn of $j_c^{mag}$ seen in Figure 4c) and the reduction of anisotropy is rooted in the multi-gap superconductivity. Interestingly, the dissipation on the 10 μΩ level, used to define H*, is weakly dependent of the current direction even at higher fields, suggesting that both "$j_c$" may continue this behavior to much higher fields.

Above liquid helium temperatures, differences between $j_c^{mag}$ and $j_c^{trans}$ become noticeable: While $j_c^{trans}$ remains high, $j_c^{mag}$ decreases and anisotropy develops. Such a difference between the $j_c$'s associated with different dissipation levels is expected from the two measurement methods and reflects the 'broadening' of the I-V characteristics at voltage levels undetectable in the present transport measurement. The length of the nanobridges (limited by the thickness of the crystals) and the sensitivity of the pulsed current experiment sets a lower bound of 50 mV/cm for the electric field criterion used to define $j_c^{trans}$, which is noticeably higher than the criteria used in industry ($\sim 1 \mu$V/cm). On the other hand, the electric fields associated with $j_c^{mag}$ are orders of magnitude lower than the ones used in transport, as this technique senses the persistent shielding currents in the sample[24]. However, this difference yields additional information about the I-V characteristic over many orders of magnitude: a broad I-V curve results in a difference between $j_c^{trans}$ and $j_c^{mag}$, while a sharp I-V characteristic gives similar results. As the difference between the two methods gradually diminishes upon cooling, the data given in Figure 4 indicates a pronounced sharpening of the I-V characteristic at low temperature. This sharpening is also directly observed in the transport I-V measurements (Inset Figure 4). A similar dependence of the critical current density on the electric field criterion has also been found in cuprates in the context of the non-monotonic field dependence of the "fish-tail" effect in magnetization[29]. As different superconducting power applications tolerate different electric fields (e.g. superconducting high power AC applications vs. DC), these details of the I-V characteristic give valuable insight on the potential applications of SmFeAs(O,F) in various devices.

We have mapped out a large region of the $j_c$(T,H) phase diagram of SmFeAs(O,F) summarized in Figure 5. H* delimits the region of onset of resistance on a scale of $10^{-4}$ $\rho_n$, while the critical current density for $j_c^{trans} \| ab$, H$\| c$ is mapped out in great detail from transport measurements in dc-fields. Furthermore the magnetically measured $j_c^{mag}$ reveals a pronounced peak that broadens and moves to higher fields at low temperatures originating from the fishtail-effect in the magnetization loop. This suggests collective pinning to become more effective at low temperatures as reflected in the convergence of the $j_c^{trans}$ and $j_c^{mag}$ values. The pinning properties have been systematically studied in the series of SmFeAs(O,F) and NdFeAs(O,F) single crystals with different F substitution levels, confirming that the fishtail effect is a general feature of these compounds. The pinning force has been analyzed by a scaling procedure providing evidence that pinning centers of only one type are responsible for the observed fishtail effect (see appendix D). For temperatures above 7 K, the peak in $j_c^{mag}(H)$ occurs in magnetic fields $H_{peak}$ available in our experiments. Extrapolating the measured peak field to lower temperatures, we estimate $\mu_0 H_{peak} \approx 10$ T at 2 K. This large value, together with the measured critical currents, places SmFeAs(O,F) in the small group of high-$T_c$ superconductors promising for applications.

In addition to the presented highly promising intra-grain transport properties, grain boundary current suppression and metallurgical issues will have to be addressed. From a microscopic physics point of view, it will be an outstanding challenge to understand how multi-band, multi-gap superconductivity with pronounced temperature dependences, e.g. of the superfluid density, will influence vortex physics and its implications for the technical critical current density.

### Acknowledgements

We thank Sergiy Katrych for performing X-Ray analyses supporting this study. Work at NHMFL-LANL is performed under the auspices of the National Science Foundation, Department of Energy and State of Florida. Electron Microscopy and FIB work was performed at the Electron Microscopy ETH Zurich (EMEZ). This work has been supported by the Swiss National Science Foundation NCCR Materials with Novel Electronic Properties (MaNEP) and by the Polish Ministry of Science and Higher Education under the research project No. N N202 4132 33.

### Author Statement

P.J.W.M and B.B. designed the experiment and wrote the paper. P.J.W.M performed the direct transport critical current experiments, the pulsed field magnetotransport was measured by P.J.W.M and F.B. R.P., K.R. and B.B. measured magnetization and evaluated the magnetic critical currents $j_c^{mag}$. K.R. performed the pinning analysis. N.Z. and J.K grew the crystals.

**Supplementary Information** accompanies the paper online.

# Figures:

**Figure 1: Four-probe resistance bars for simultaneous *c*-axis and *ab*-plane resistivity measurements carved out of a SmFeAsO$_{0.7}$F$_{0.25}$ single crystal using the Focused Ion Beam (FIB)**

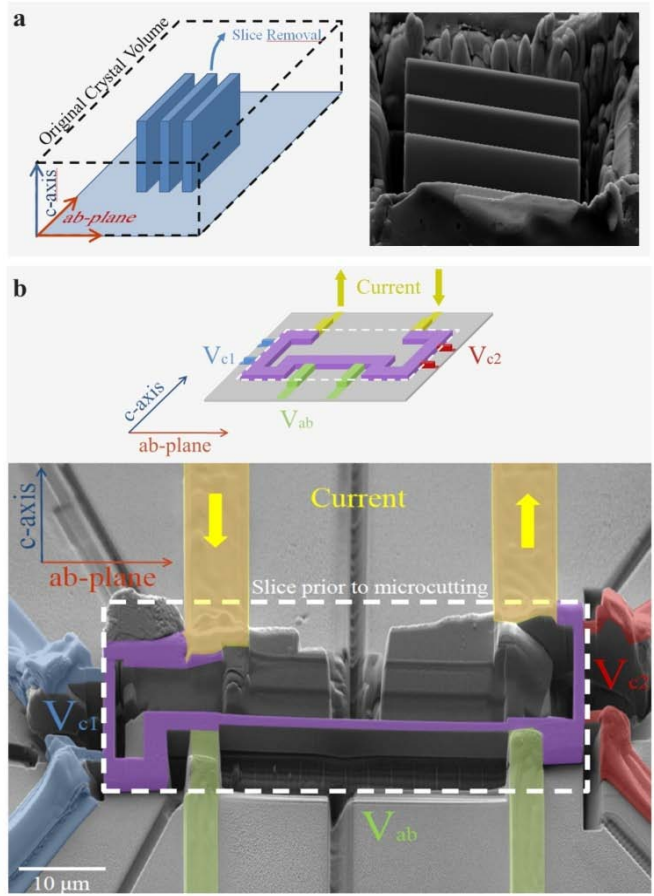

a) A crystal is positioned on a substrate with the *c*-axis pointing perpendicular to the plane. The dashed volume indicates the original crystal that is removed during FIB cutting, leaving only the lamellae standing.

b) The lamella is transferred to another substrate and flipped, so that its *c*-axis is now aligned in the plane (short edge). Most of this lamella (dashed line) is again removed leaving only the small current path standing (violet). Eight platinum leads are deposited onto the crystal edges (all other colors) that are connected to the resistance bars by narrow (~800 nm for *c*-axis) crystal bridges. The common current is injected through the yellow contacts and traverses two *c*-axis resistance bars (blue & red voltage contacts) and one along the *ab*-plane (green voltage contacts). Dimensions of resistance bars: length ~ 35 μm (*ab*-plane), ~ 5 μm (*c*-axis), cross-section ~1.5 μm$^2$

**Figure 2: Normal State Resistivity Ratio of SmFeAsO$_{0.7}$F$_{0.25}$**

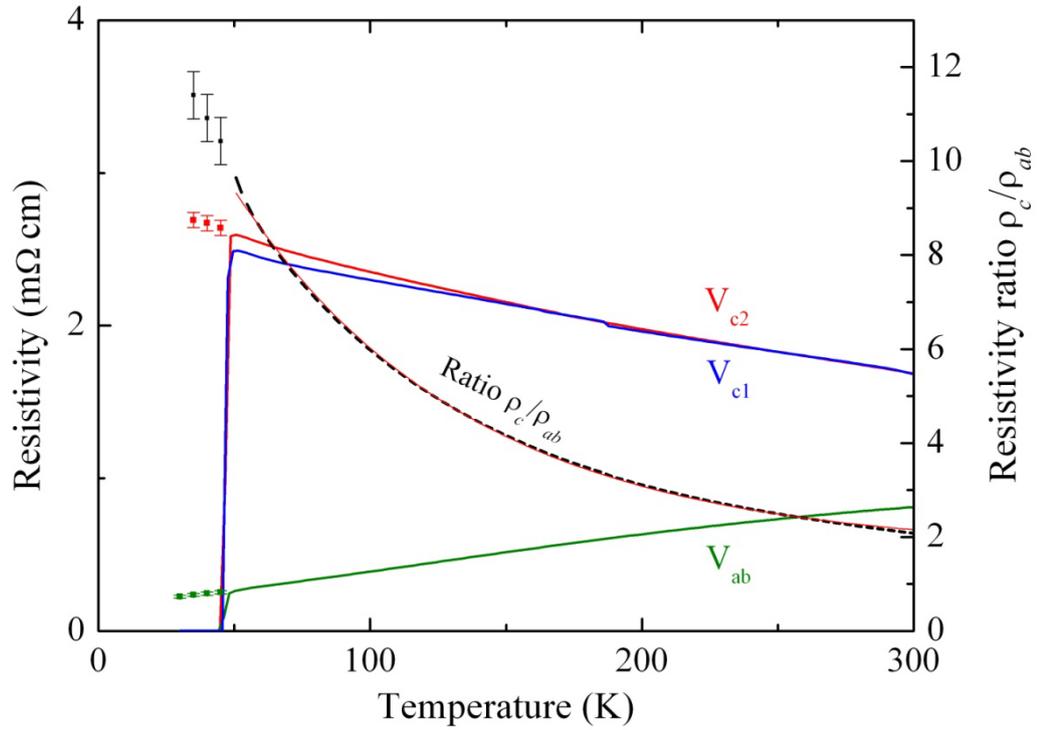

The normal state resistivity $\rho_c$ along the *c*-axis of SmFeAsO$_{0.7}$F$_{0.25}$ in zero-field increases with decreasing temperature, while the *ab*-plane resistivity $\rho_{ab}$ decreases. The resulting resistivity ratio $\rho_c$ / $\rho_{ab}$ (dashed line) is fitted very well by an exponential increase exp(-T/T$_0$) + 1.73, with T$_0$ = 86.4 ($\pm$0.6) K (red line). Similar behavior was observed in different samples.

**Figure 3: Magnetoresistance of SmFeAsO$_{0.7}$F$_{0.25}$ in pulsed fields up to 65 T at various temperatures for fields and currents along and perpendicular to the *c*-axis**

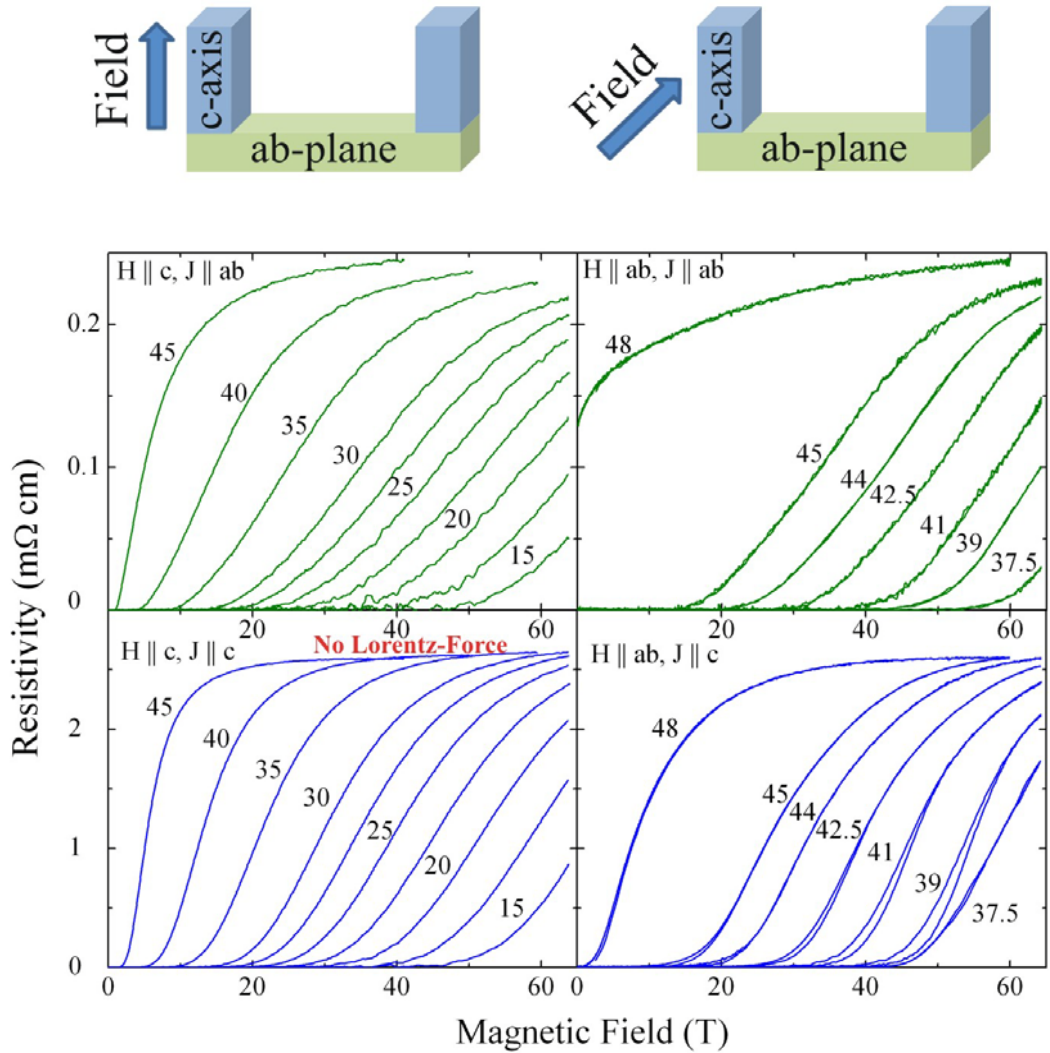

The four panels correspond to the various combinations of currents and fields oriented along or perpendicular to the crystallographic *c*-axis. For the lowest temperatures with j||*c*, a small hysteresis is observed if the field is applied in the FeAs planes. The hysteretic width depends on the field sweep rate and vanishes at low rates, indicating irreversible vortex trapping.

**Figure 4 : Critical Current Density of SmFeAsO$_{0.7}$F$_{0.25}$**

a)

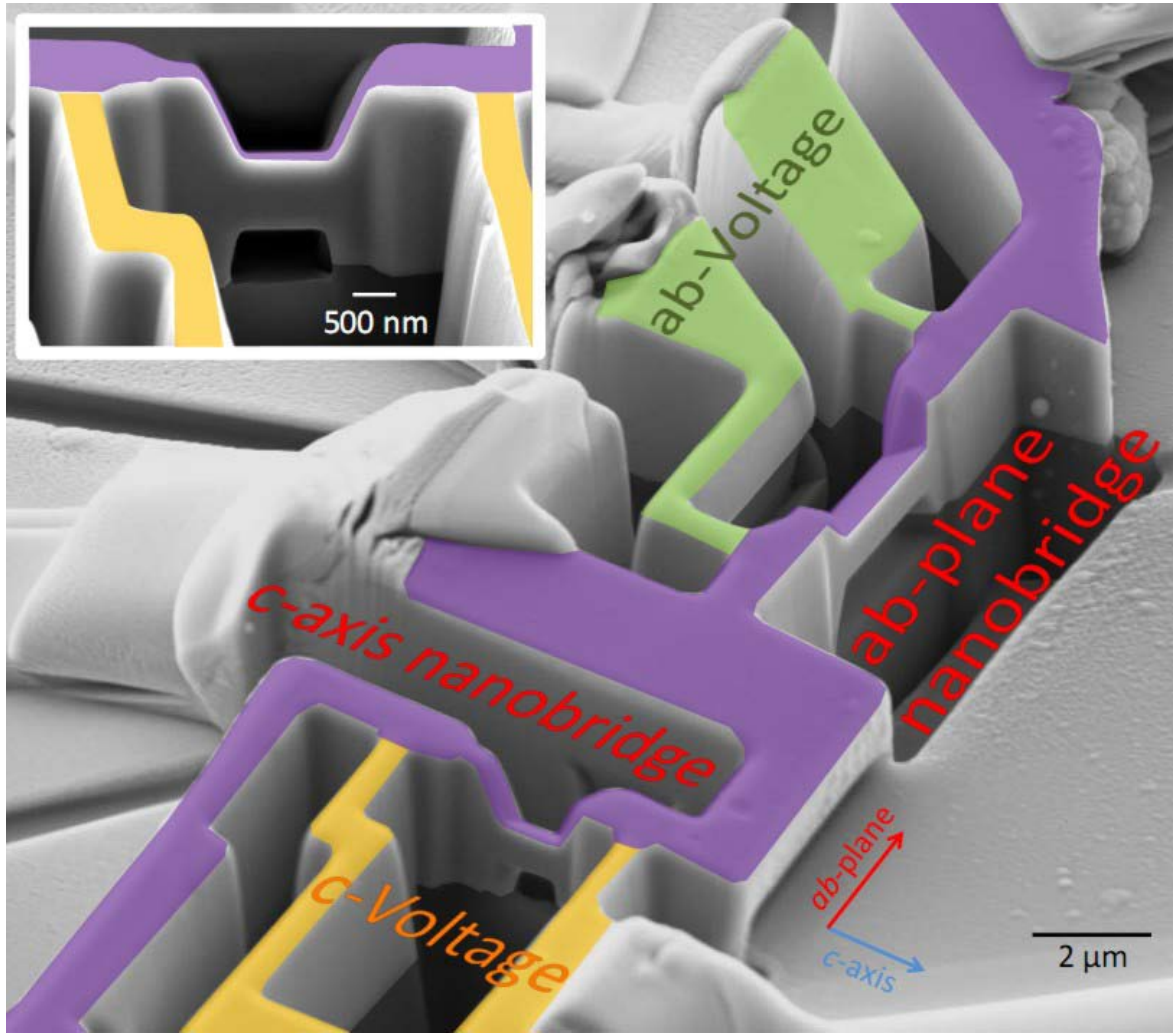

b)

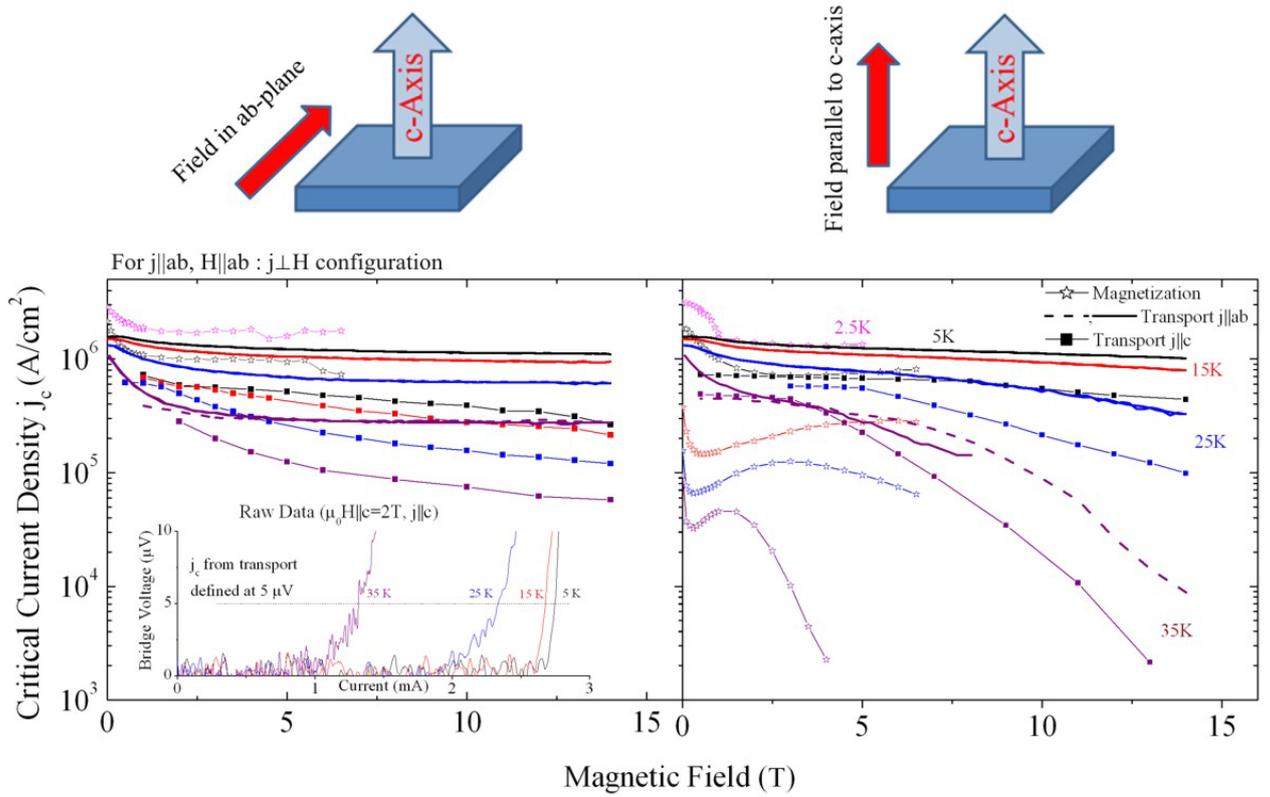

c)

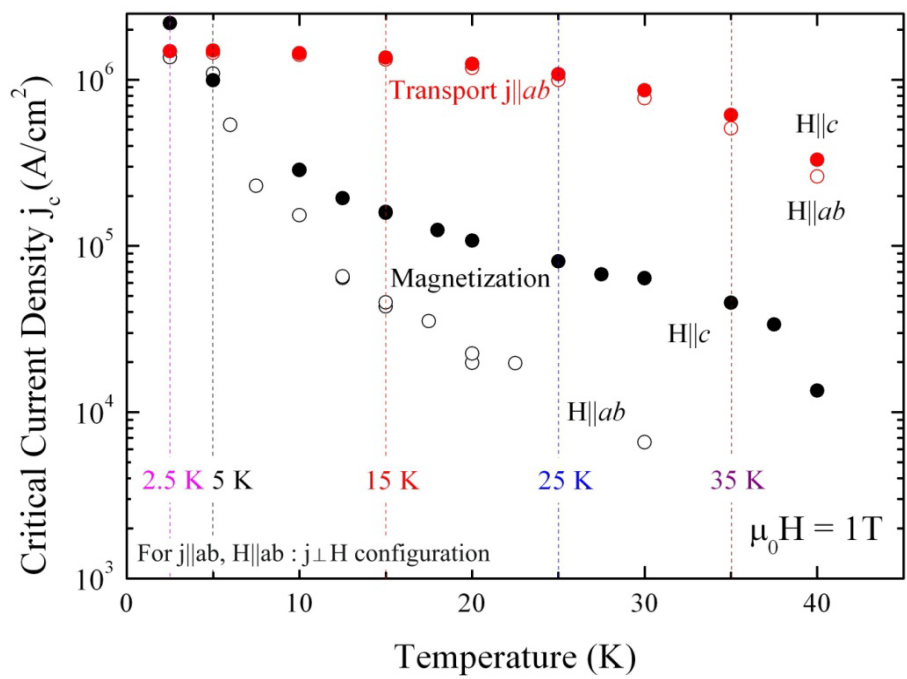

a)

Two free-standing $SmFeAsO_{0.7}F_{0.25}$ nanobridges (cross-section ~ 600 nm x 600 nm, length of narrow part ~ 1 - 3 μm) cut into a single crystal lamella (similar to Fig. 1). Due to the special cutting procedure, the *c*-axis is in the plane of measurement, and this structure allows the direct measurement of $j_c$ along different axes. Three contacts for current injection were made on the sides of the crystal to ensure the current passes one bridge only during a measurement cycle.

b)

Critical current density $j_c^{mag}$ of $SmFeAsO_{0.7}F_{0.25}$ determined from magnetic hysteresis loops (stars) after Bean[24] and $j_c^{trans}$ from direct critical transport experiments across the nanobridge shown in a) as well as another, similar sample from the same growth batch. Inset: Raw voltage data during pulsed current ramp.

c)

Temperature dependence of $j_c\|ab$ for $\mu_0 H = 1T$. At low temperatures, the critical current anisotropy $j_c(H\|ab)/j_c(H\|c)$ is reduced and the values of $j_c$ converge.

**Figure 5: Region for potential application of SmFeAs(O,F)**

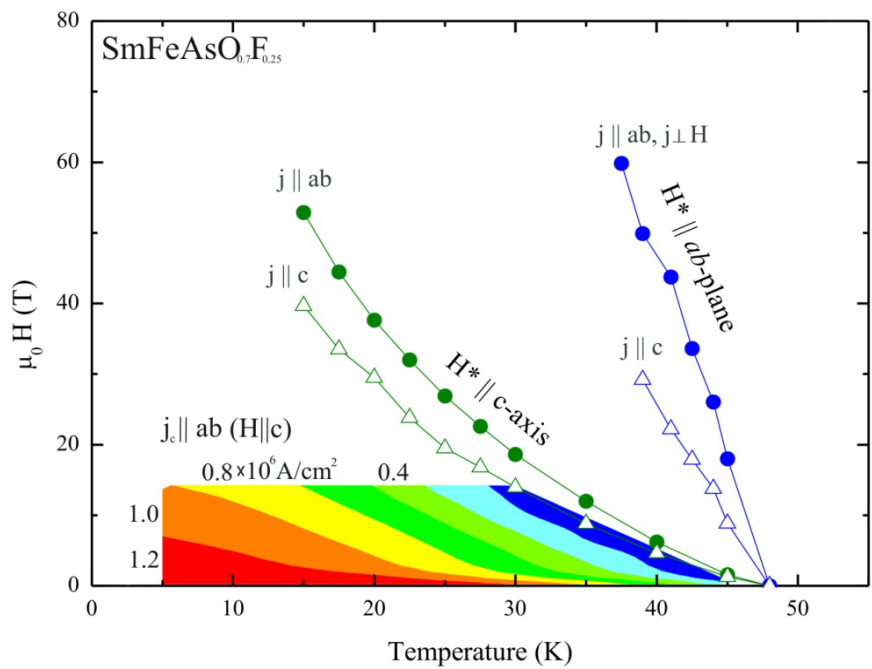

A color map of the critical current density $j_c^{trans}\|ab$ (H$\|$c) in SmFeAsO$_{0.7}$F$_{0.25}$ bordered by H* (marking the dissipation level of 10 $\mu\Omega$cm) measured in pulsed fields. The technically limiting H*$\|c$ (j$\|c$) reaches 40 T at 15 K. H*$\|c$ shows no signs of saturation and its slope of 3.4 T/K at 15 K suggests a significant further increase down to 4.1 K. The measurement j$\|ab$, H$\|ab$ was performed in a full Lorentz force configuration, as it is the technologically most important. The critical current density increases steeply with decreasing temperature at any given field and reaches a high and only weakly field dependent value at low temperatures. These favorable intra-grain properties suggest a large H-T region interesting for application.

# Appendix

**Appendix A: "Upper critical field $H_{c2}$" derived from resistive transitions**

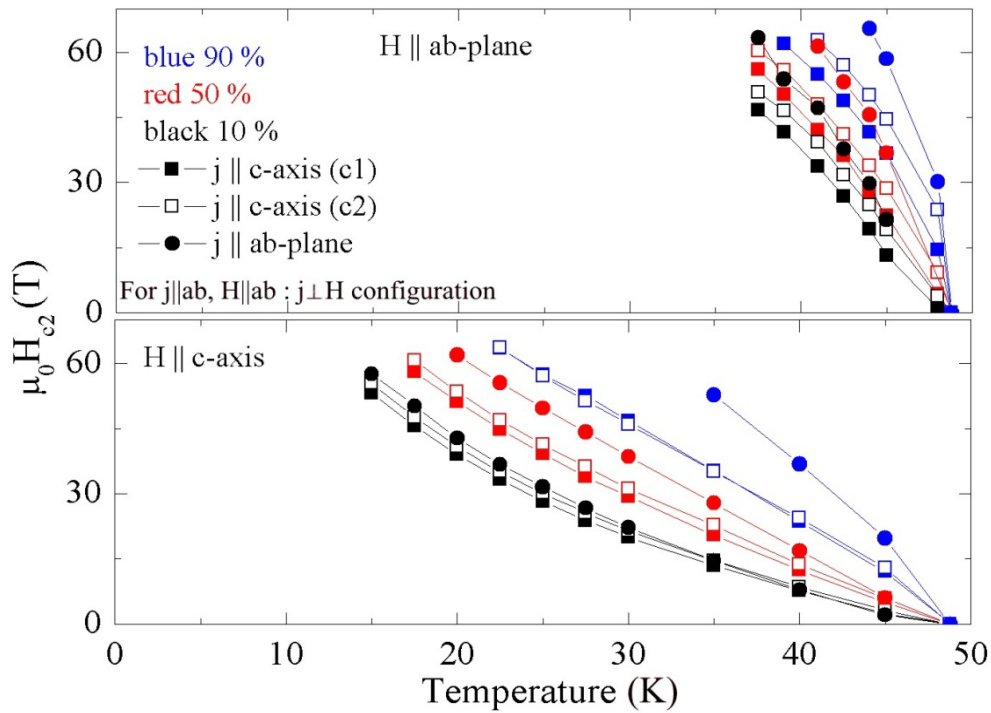

**Figure A1:** The normal state resistivity $\rho_n$ was extrapolated into the superconducting state and the magnetic field H at which $\rho(T,H) = x \, \rho_n$ (x=10%, 50%, 90%) is called '$H_{c2}$'. While the 50% criterion '$H_{c2}$' is widely accepted to be close to the real $H_{c2}$, the large dependence of '$H_{c2}$' on the arbitrary choice of criteria prevents an unambiguous definition. The onset of the transition (10%) does not depend on the current direction, but the transition is broader (Fig. 3) for current in the *ab*-plane, leading to a separation of the 'upper critical fields' for the two current directions. J ∥ *ab* is similar compared to the reported behavior in NdFeAs(O,F) [15] in low fields, but does not show a comparably pronounced upturn for all criteria. The 50% '$H_{c2}$' is almost linear up to 65 T, and the 90% '$H_{c2}$' even shows a downturn, which indicates that the transition may become sharper at lower temperatures and the material more isotropic.

**Appendix B: Temperature dependence of the H$_{c2}$ anisotropy**

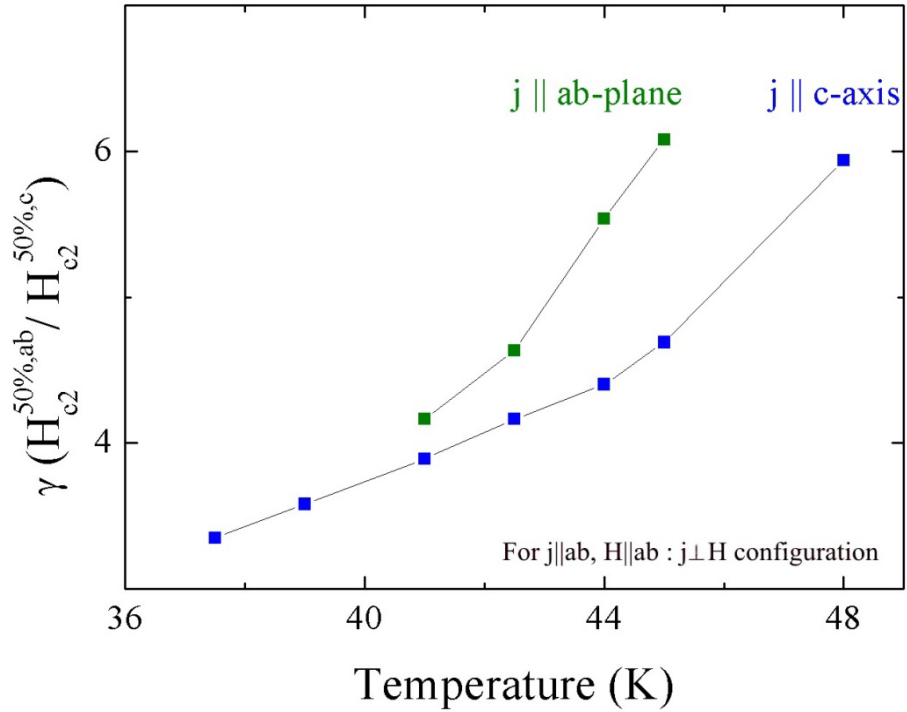

**Figure B1:** Temperature dependence of the '50% H$_{c2}$' (Appendix A) anisotropy parameter $\gamma$(T) = H$_{c2}$∥*ab* (T) / H$_{c2}$∥*c* (T) for currents along the *c*-axis and in the *ab*-plane. The anisotropy decreases with decreasing temperature from 6 to 3.4, caused by the downturn of 'H$_{c2}$'∥*ab*. This behavior is common in the pnictides and was reported in '1111'[15,7] and '122'[18] compounds.

**Appendix C: Comparison of various SmFeAsO$_{0.7}$F$_{0.25}$ nanobridges**

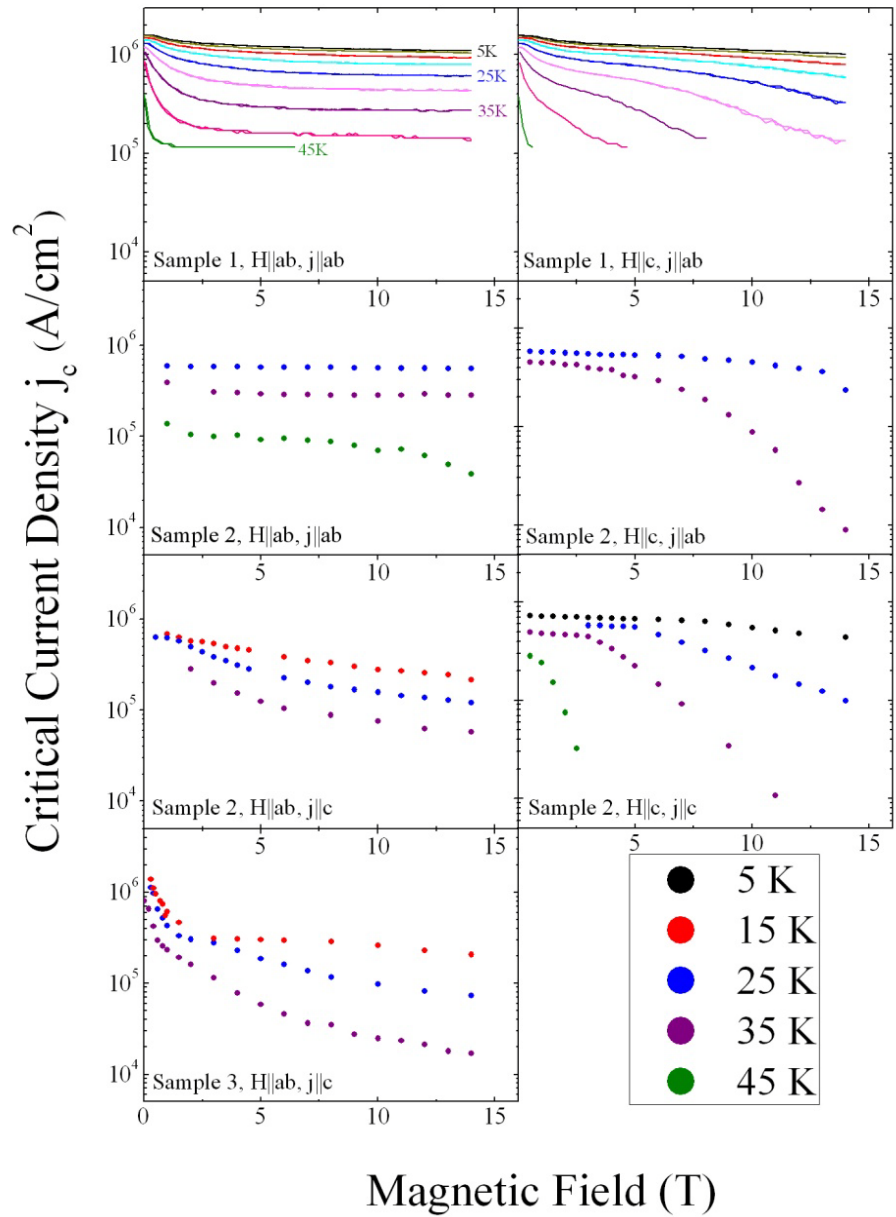

**Figure C1:** We have performed critical current measurements on various different nanobridges employing two different transport measurements, using both pulsed (dots)

and linearly ramped (lines) currents, and confirmed our main results of high, nearly isotropic and only weakly field dependent critical current densities in every one of them. All H∥ab, j∥ab experiments were performed in the Lorentz force (j⊥H) as it is the technologically more important configuration.

During the quench of the nanobridge a considerable amount of energy is deposited and some mechanically weak samples burned out during measurement. This was confirmed by SEM investigations of these damaged bridges, which showed cracks. Therefore, not all temperatures could be measured for all samples.

## Appendix D: Peak Effect Scaling

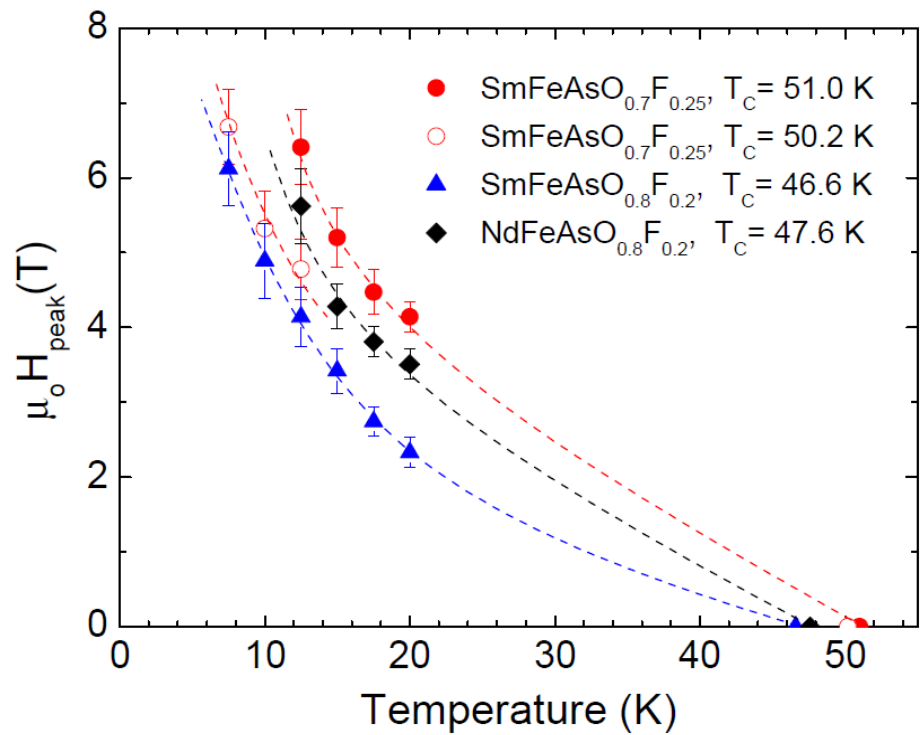

**Figure D1:** Characteristic field $H_{peak}$ where the maximum of magnetically measured $j_c$ occurs for the fish-tail shaped $j_c(H)$ dependence, for series of SmFeAs(O,F) single crystals with a different F substitution level and therefore different $T_c$'s, together with results for a NdFeAs(O,F) single crystal for comparison. The magnetization loops used for $j_c(H)$ calculations were obtained for H oriented parallel to the *c*-axis, thus for currents flowing in the favourable *ab*-plane, as in the case of most applications. Extrapolating the measured peak field to lower temperatures, where it exceeds the range accessible in our measurements, we estimate $\mu_0 H_{peak} \approx 10$ T at 2 K. This large value, together with the measured critical currents, seems to be promising for applications. The crystals were too small to be measured for the H∥*ab*-plane configuration.

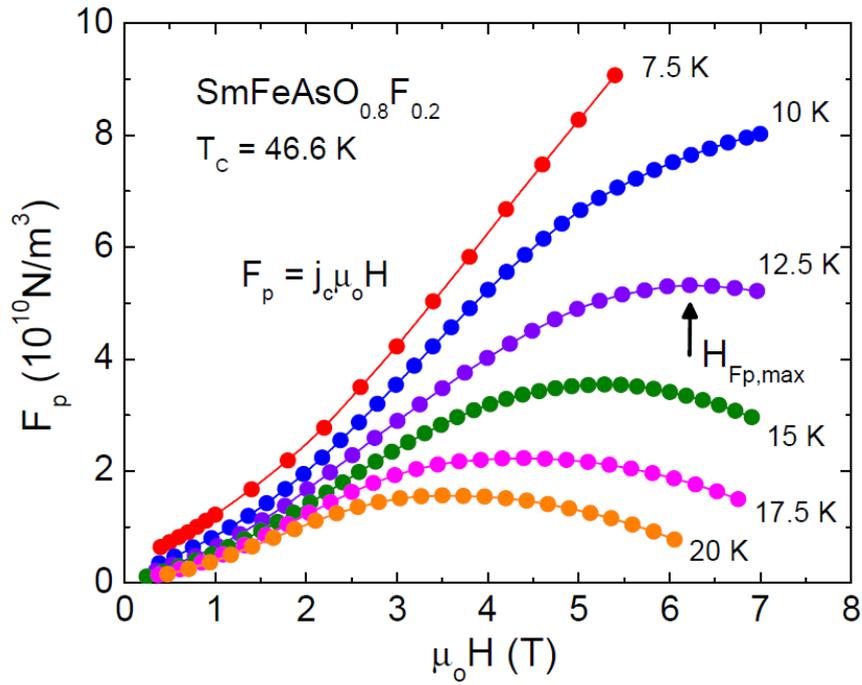

**Figure D2:** Volume pinning force $F_p = j_c\mu_0 H$ for the Sm1111 single crystal with the highest magnetically measured critical currents (e.g., $\approx 1.3 \times 10^6$ A/cm$^2$ at 10 K and 5 T), obtained for H parallel to the *c*-axis. For temperatures above 10 K, $F_p(H)$ reaches a maximum, $F_{p,max}$, in a magnetic field $H_{Fp,max}$ accessible in our experiments. The pinning force can be now analysed by a scaling procedure as explained in the next Figure.

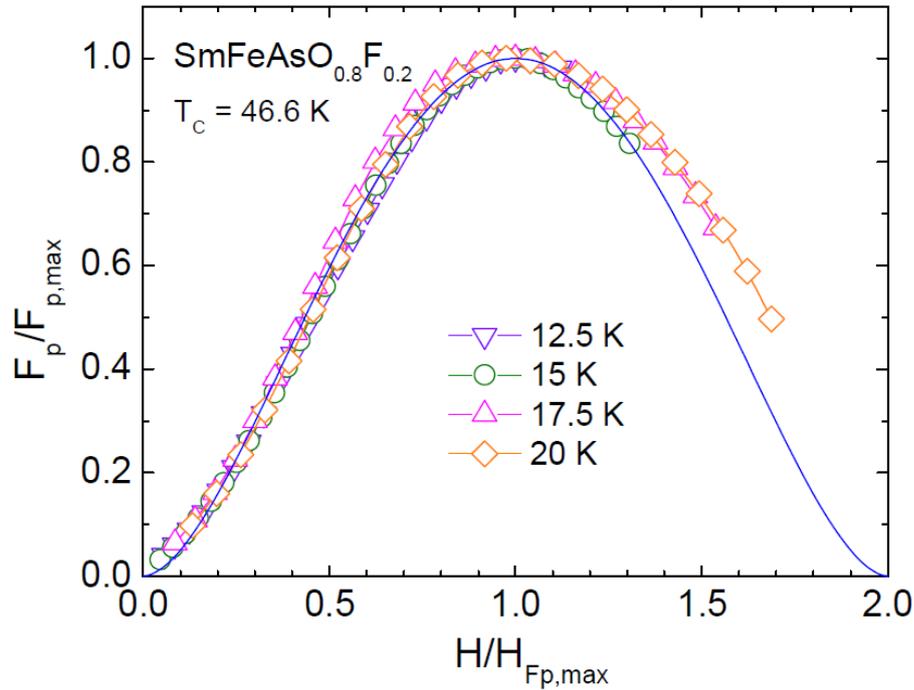

**Figure D3:** Reduced pinning force $F_p/F_{p,max}$ versus reduced magnetic field $H/H_{Fp,max}$ plotted for data shown in Fig.D2. The pinning force scales very well, for all temperatures at which a maximum in the $F_p(H)$ dependence is observed. This analysis of pinning bases on the Dew-Hughes[30] model with a modified scaling law f(h) = h$^p$(2-h)$^q$, where f = $F_p/F_{p,max}$, h = H/H*, H* is a scaling field, and *p* and *q* depend on the relevant pinning mechanism. This model and its numerous modifications were successfully used to analyze pinning properties in several high-$T_c$ superconductors.[31,32] As a scaling field H* we use $H_{Fp,max}$ (see Fig. D2) and fit the experimental results well for *p = q = 1.8* (blue line). The scaled curves in the temperature range from 12.5 to 20 K collapse to a single curve, suggesting that a single pinning mechanism is responsible for the increase of critical currents in the "fishtail" region. A small deviation from the scaling law may be caused by the flux creep that intensifies at higher temperatures. Similar scaling has been observed for all single crystals of the "1111" family we have studied indicating that the strong intrinsic pinning centres of a similar type dominate in this material in the explored temperature and field regions. The possible pinning centers may originate in the local phase variation due to the oxygen or/and fluorine inhomogeneities. Above 20 K the SQUID signal was ~ 2 x 10$^{-8}$ emu, thus too small to be measured.

**Appendix E:** Preparation of *c*-axis transport sample (U-shape)

After confirming the crystal quality by magnetic and X-Ray characterization, a thin (~ 1 µm) free standing lamella was carved out of the crystal with the FIB (Fig. 1a). In a second step, this lamella was removed from the remaining crystal and laid flat on a $SiO_2$ substrate. In this new configuration, the c-axis of the crystal is now in the substrate plane and accessible for contacting by ion-assisted platinum deposition on the edges. For an ideal four-point measurement, a long bar with small cross-section and pointlike voltage contacts attached to the side is desirable, and this requirement can be achieved via FIB. In a third step, a U-shaped structure is defined with two of the legs along the *c*-direction and one along the *ab*-plane. On each leg two narrow side protrusions (~ 800 – 900 nm) are left during cutting which serve as optimal, low resistance side contacts for voltage measurements (Fig. 1b). Similar lamellae are used to prepare the critical current nanobridges (Fig. 4a), which involves more sophisticated FIB procedures.